# Frequency-Temperature sensitivity reduction with optimized microwave Bragg resonators

J-M. Le Floch,[1, 2, a)] C. Murphy,[2] J.G. Hartnett,[3] V. Madrangeas,[4] J. Krupka,[5] D. Cros,[4] and M.E. Tobar[2]

[1)]*MOE Key Laboratory of Fundamental Physical Quantities Measurement, School of Physics, Huazhong University of Science and Technology, Wuhan 430074, Hubei, China*

[2)]*School of Physics, The University of Western Australia, Crawley, Western Australia 6009, Australia*

[3)]*Institute for Photonics and Advanced Sensing (IPAS) and the School of Physical Sciences, University of Adelaide, Adelaide, S.A. 5005, Australia*

[4)]*XLIM, UMR CNRS 7252, Université de Limoges, 123 av. A. Thomas, 87060 Limoges Cedex, France*

[5)]*Instytut Mikroelektroniki i Optoelektroniki PW, Koszykowa 75, 00-662 Warsaw, Poland*

(Dated: 30 December 2016)

# Frequency-Temperature sensitivity reduction with optimized microwave Bragg resonators

Dielectric resonators are employed to build state-of-the-art low-noise and high-stability oscillators operating at room and cryogenic temperatures. A resonator temperature coefficient of frequency is one criterion of performance. This paper reports on predictions and measurements of this temperature coefficient of frequency for three types of cylindrically-symmetric Bragg resonators operated at microwave frequencies. At room temperature, microwave Bragg resonators have the best potential to reach extremely high Q-factors. Research has been conducted over the last decade on modeling, optimizing and realizing such high Q-factor devices for applications such as filtering, sensing, and frequency metrology. We present an optimized design, which has a temperature sensitivity 2 to 4 times less than current whispering gallery mode resonators without using temperature compensating techniques and about 30% less than other existing Bragg resonators. Also, the performance of a new generation single-layered Bragg resonators, based on a hybrid-Bragg-mode, is reported with a sensitivity of about -12ppm/K at 295K. For a single reflector resonator, it achieves a similar level of performance as a double-Bragg-reflector resonator but with a more compact structure and performs six times better than whispering-gallery-mode resonators. The hybrid resonator promises to deliver a new generation of high-sensitivity sensors and high-stability room-temperature oscillators.

a)Electronic mail: jm_lefloch@hust.edu.cn

Frequency-Temperature sensitivity reduction with optimized microwave Bragg resonators

# I.  INTRODUCTION

Dielectric resonators were initially introduced into microwave technology to increase the performance and to reduce the size of filters and resonators. In general, such resonators may operate in a variety of different electromagnetic modes[1], depending on their intrinsic losses, targeted field confinement, and resonance frequencies. For decades, simple dielectric resonators have been the best choice due to their robustness in harsh environments and low cost. More specifically, in the domain of frequency metrology and precision measurements, the use of high Q-factor, and low-loss single-crystal sapphire dielectric resonators, even though more costly, have had great success in narrow band filters[1] and state-of-the-art oscillators both at room[2-4] and cryogenic temperatures[5-13]. Other more recent applications include characterizing bulk and thin film materials[14-17], as well as being used as ultra-high sensitive sensors[18].

At room temperature and X-band frequencies, whispering-gallery-mode (WGM) resonators[3,19,20] have Q-factors of about 100,000 to 200,000 for modes with dominant magnetic (denoted WGE) and electric (denoted WGH) polarizations parallel to the cylindrical crystal axis, respectively[3,20,21]. Additionally, the temperature coefficient of frequency for WGE and WGH modes have been measured to be about -50 and -70 ppm/K, respectively[22]. Because of this large sensitivity, other types of electromagnetic modes have to be employed in an effort to overcome this. Such a choice might be the transverse electric modes[23,24] but they do not exhibit as high Q-factors as WGM, or photonic band gap[25-29] and Bragg effect resonators[30-37], which involve bigger volumes. In the millimeter wave frequency band, dimensions of the dielectric resonators become very tiny[38,39], thus their sensitivity to temperature changes increases[40]. To solve this problem, the use of photonic band gap (PBG) resonators offers a scaling factor which allows the design of larger resonators whereas Bragg resonators cannot be as big. However, at X-band (6-12GHz), PBG are too large. Hence, Bragg resonators remain a viable option. The latter has been predicted to reach a Q-factor of 1 million at 10GHz[38,41] with multiple layers, corresponding to a factor five times better than WGM resonators[3,4]. This Q-factor improvement between both electromagnetic modes comes from the mode distribution of the field inside the cavity. WG modes confine the electric energy into the sapphire whereas the Bragg mode confine it in its inner free-space region and has little energy

Frequency-Temperature sensitivity reduction with optimized microwave Bragg resonators in its dielectric reflectors. Thus, Bragg resonators are less limited by the material of its reflector[1,41,42]. In terms of volume between both resonators with same frequency, Bragg resonators suffer of larger volume compared to WG mode resonators[41]. The size of Bragg resonators is set by the height of the inner free-space region which corresponds to half of a wavelength. This dimension in free-space is then bigger than in a dielectric.

A Bragg effect resonator consists of multiple layers of different dielectric materials, and it enables the confinement of the electromagnetic field to the center of the resonator. This is due to the destructive interference in the outer layers of the resonator and constructive interference in the center. The center of the cavity consists of either vacuum or low-loss material[42]. The field confinement in the inner free-space region of the Bragg resonator reduces the effect of the surface resistance of the metal enclosure and increases the geometric factor of the cavity. It means the Bragg resonator design can enhance the unloaded Q-factor of a resonator producing an unloaded Q-factor as high as ten times the dielectric loss limit[43]. For example, an optimized Sapphire distributed Bragg resonator has a Q-factor 1.5 times bigger than any WGM[41], then double-reflector structures, increases the difference to 3.5 times to finally reach the highest limit with a factor 5 using triple-reflector resonator. Spherically and cylindrically-symmetric Bragg reflector resonators have successfully achieved this milestone[44–47].

Oscillators involved in precision measurement experiments require high-Q factor resonators for low-phase noise and high-frequency stability at microwave and millimeter wave frequencies. The short-term frequency stability degradation is the result of the frequency drifting more than the resonator bandwidth, which reduces with increased Q-factor. First, it depends on their resonance frequency varying with temperature, resulting in a change of dimension and material properties (coefficient of thermal expansion (CTE) and temperature coefficient of permittivity (TCP)). The TCP corresponds to the rate of change of the dielectric permittivity with respect to temperature. The resonator frequency to temperature dependence is denoted by the temperature coefficient of frequency ($\tau_f$) and it can be measured quantitatively. It quantifies the rate of change of the resonance frequency with respect to temperature and is measured by stepping the resonator temperature and recording the resulting resonance frequency when it has stabilized. This sensitivity to temperature requires high constraints on the frequency stabilization electronics of the oscillator and thermoelectric cooler feedback to maintain the center frequency of the resonator, which determines the oscillator frequency. The same applies for material characterization and sensor applications, where a frequency shift determines the right material properties or the detection of an event in case of a sensor. A perfect thermally stable resonator would have a temperature coefficient of zero. Nevertheless, one can predict the frequency to temperature sensitivity from dielectric properties and simulation with rigorous electromagnetic simulation software. In our case, we employ the method of lines and finite element analysis[41,46]. For instance, we used sapphire ($Al_2O_3$) from GTAT Crystal Systems. It is a uniaxial anisotropic material. Therefore, the coefficient of thermal expansion (CTE) and temperature coefficient of permittivity (TCP) of sapphire have different values along the perpendicular and parallel directions to the crystal axis. The successful design of frequency-temperature stable resonant structures can be obtained either with the contraction (CTE) compensating the material TCP[35] or with a compensation technique employing two materials with TCPs of opposite sign[33]. Thus, to construct a compensated resonator; it is very critical to know the different temperature dependencies of materials (TCP, CTE)[22,42].

In this paper, we report a model extension of the $\tau_f$ coefficient prediction for a two-dimensional cylindrically-symmetric Bragg resonator and the measurements of a single sapphire reflector, a single alumina reflector, and a double-sapphire-alumina as well as for a hybrid-Bragg-mode within the same single alumina reflector resonator[43].

## II. BRAGG RESONATOR DESIGN

Conventional Bragg resonators[30–37,46,48] enable the confinement of the electromagnetic field within an inner free-space region with the help of partial mirrors (reflectors) made of a pair of dielectrics and air layers. The height of the inner free-space region sets the resonance frequency, and the thickness of the dielectric leads to destructive interference in such a way that the field confinement in the inner region increases. For both spherical and cylindrical symmetries, the thickness of the dielectric is about a quarter of the guided wavelength (see Eq.1).

# Frequency-Temperature sensitivity reduction with optimized microwave Bragg resonators

$$\lambda_g = \frac{c}{f\sqrt{\epsilon_{eff}}}, \qquad (1)$$

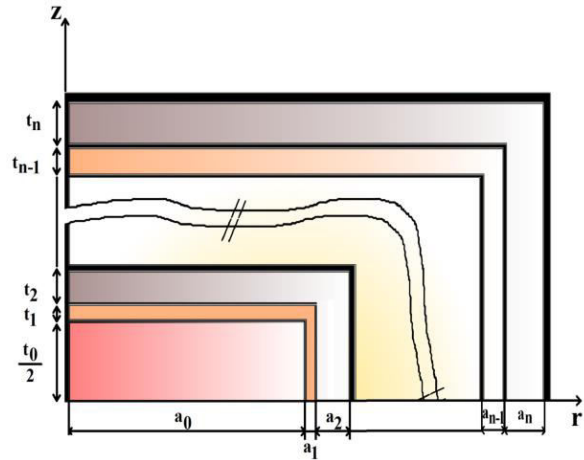

FIG. 1: Schematic of a Bragg resonator with *N*-reflectors. A reflector in either directions (radial and/or axial), is composed of one layer of dielectric and one layer of free-space. In our case, we have Bragg resonator designs with the same number of radial and axial reflectors. The inner region defined by $a_0$ and $t_0/2$, locates where the electromagnetic field is maximum.

where $\lambda_g$ is the guided wavelength, $c$ is the light velocity, $f$ the resonance frequency and $\varepsilon_{eff}$ is the effective permittivity.

In recent years, new concepts of cylindrically-symmetric Bragg resonators have been modeled and demonstrated[41–43,47]. These exhibit a more compact structure, a more confined energy in the inner region of the resonator, and thus, a much higher Q-factor. We demonstrated that a linear combination of electromagnetic modes occurs in the horizon- tal reflectors. We created a non-Maxwellian factor ($\gamma$), to correct this non-desired effect of mode combination. This determination then allowed the calculation of the resonator dimensions. The modeling was completed first for a particular case[41], and then, ex- tended to a design with an arbitrary reflector thickness[47] to generalize and enhance the capabilities of Bragg resonators. Using a separation of variables technique, two sets of independent equations were developed for both directions, to simultaneously determine all the required resonator dimensions (see the decomposition of the parameters on Fig.1). We assume the field propagates as sine function along z-axis and as a Bessel function $J'_n$
along r-axis. Along the axial direction, the different thicknesses can be expressed as:

Frequency-Temperature sensitivity reduction with optimized microwave Bragg resonators

$$t_{2i} = \frac{\pi(1-\eta_i)(2s_i-1)}{k_0}, \quad (2a)$$

$$t_{2i-1} = \frac{\pi\eta_i(2s_i-1)}{k_0\sqrt{\epsilon_i}}, \quad (2b)$$

$$t_0 = \frac{\gamma\pi p}{k_0}, \quad (2c)$$

for $i = 1$ to $N$, where $N$ defines the number of pairs of Bragg reflector layers, consisting of one layer of dielectric and a layer of air. $\eta_i$ corresponds to the proportion of dielectric in the $i^{th}$ reflector. $s_i$ the number of mode variation within the $i^{th}$ reflector in the axial direction. Here, $\gamma$ is the non-Maxwellian parameter allowing the correction of the dielectric layer thickness accordingly to the mode distribution and frequency. In the radial direction, the set of equations is:

$$a_{2i} = \frac{\gamma(1-\rho_i)}{k_0\sqrt{\gamma^2-1}}, \quad (3a)$$

$$a_{2i-1} = \frac{\gamma\rho_i}{k_0\sqrt{\epsilon_i\gamma^2-1}}, \quad (3b)$$

$$a_0 = \frac{\gamma\chi_{1n}}{k_0\sqrt{\gamma^2-1}}, \quad (3c)$$

for $i = 1$ to $N$, where $N$ defines the number of pairs of Bragg reflector layers, consisting of one layer of dielectric and a layer of air. $\rho_i$ corresponds to the proportion of dielectric in the $i^{th}$ reflector. $q_i$ the number of mode variation within the $i^{th}$ reflector in the radial direction. $\gamma$ is determined from the aspect ratio of the resonator height over its radius. This is defined as follow:

$$AR_{n,m,p}^{nrr,nra}(\gamma,\epsilon_r) = \frac{\frac{\pi p \sqrt{\gamma^2-1}}{2\chi_{1,n}}(1+\sum_{j=1}^{nra}\frac{2(2s_j-1)}{\gamma p}(1-\eta_j+\frac{\eta_j}{\sqrt{\epsilon_j}}))}{1+\frac{1}{\chi_{1n}}\sum_{j=1}^{nrr}(1-\rho_j)(\chi_{1,j+n+2q_j-2}-\chi_{1,j+n-1})+\rho_j\sqrt{\frac{\gamma^2-1}{\epsilon_j\gamma^2-1}}(\chi_{1,j+n+2q_j-2}-\chi_{1,j+n-1})} \quad (4)$$

where $\chi$ is the root of the first order derivative of Bessel function, $m$ is the number of azimuthal variations (here, $m = 0$), $n$ represents the number of radial variations, $p$ is the number of axial variations in the inner region. $nrr$ and $nra$ are the number of variations in the radial and the axial reflectors respectively.

### III.  THE TEMPERATURE COEFFICIENT OF FREQUENCY

The temperature coefficient of frequency ($\tau_f$) can be predicted as previously demonstrated with a spherical Bragg resonator[35,38], where the set of equations given in the literature is one dimensional along the radial direction of the structure. In our case, for a cylindrical Bragg resonator, the system of equations becomes two-dimensional, related to the radial and axial directions of the structure. Thus, in this work, we necessarily extended the prediction of $\tau_f$ from spherical to cylindrical topology in the following.

If we can determine the resonance frequency fluctuations as a function of the changes in permittivity and dimensions of the resonator, then the desired coefficient $\tau_f$ can be calculated by taking the derivative of the function with respect to temperature and thus verified experimentally. To measure the temperature coefficient of frequency of a resonator, it is necessary to control the temperature and track the resonance frequency of the Bragg mode with a vector network analyzer referenced to a H-Maser. This measurement relies on S-parameter transmission ($S_{21}$) technique for better precision. Both the controller and the analyzer were remotely controlled with an in-house GPIB protocol standalone software. Once the frequency stabilizes within 10kHz, we record the value and step the temperature to the next set point. This way it gives the frequency dependence of the resonator to the temperature fluctuations. Proceeding with a linear fit around a specific chosen temperature point, the coefficient of the linear regression gives the temperature coefficient of frequency.

#### A.  Prediction

When the manufacturers quote the temperature stability of dielectrics, they usually give it for the case where the thermal expansion of the metal is negligible. Such conditions are when the dielectric resonator is in free space and resonating in either a WGM or a TE$_{01\delta}$ mode. This measurement determines $\tau_f$ and can be predicted from below: the general formula to predict the temperature coefficient of frequency, $\tau_f$ for a two-dimensional resonator has been extended from the spherical Bragg resonator[38] as follows:

# Frequency-Temperature sensitivity reduction with optimized microwave Bragg resonators

$$\tau_f = \left(\kappa_{f\epsilon_1}\tau_{\epsilon_1} + \kappa_{f\epsilon_2}\tau_{\epsilon_2} + \kappa_{f\epsilon_3}\tau_{\epsilon_3}\right) + \kappa_{fdiel}\alpha_{diel} + \kappa_{fmet}\alpha_{met} \qquad (5)$$

where $\kappa_{f\epsilon i}$ is related to the electric field confinement in the region i, $\tau_{\epsilon i}$ is the temperature coefficient of permittivity (TCP), $\kappa_{fdiel}$ the dimensional frequency coefficient of the Bragg reflector region, $\alpha_{diel}$ is the coefficient of thermal expansion of the Bragg reflector region, $\kappa_f$ met is the dimensional frequency coefficient of the metallic enclosure and $\alpha_{met}$ is the coefficient of thermal expansion of the metal. In the single Bragg resonator, Eq. 5 can be simplified with $\epsilon_{1;3} = 1$ for the permittivity of vacuum with a null TCP. We then define the $\kappa_f\ \epsilon_2$, $\kappa_f$ diel and $\kappa_f$ met as follows:

$$\kappa_{f\epsilon_i} = \frac{\partial f}{\partial \epsilon_i}\frac{\epsilon_i}{f} = -\frac{1}{2}pe_i \qquad (6a)$$

$$\kappa_{fdiel_\perp} = \left(\frac{\partial f}{\partial r_j}\frac{r_j}{f} + \frac{\partial f}{\partial r_{j+1}}\frac{r_{j+1}}{f}\right) \qquad (6b)$$

$$\kappa_{fdiel_\parallel} = \frac{1}{2}\left(\frac{\partial f}{\partial h_j}\frac{h_j}{f} + \frac{\partial f}{\partial h_{j+1}}\frac{h_{j+1}}{f}\right) \qquad (6c)$$

$$\kappa_{fdiel} = \kappa_{fdiel_\perp} + \kappa_{fdiel_\parallel} \qquad (6d)$$

$$\kappa_{fmet} = \left(\frac{\partial f}{\partial r_{j+2}}\frac{r_{j+2}}{f} + \frac{1}{2}\frac{\partial f}{\partial h_{j+2}}\frac{h_{j+2}}{f}\right) \qquad (6e)$$

where $\epsilon_i$ describes the permittivity in the different regions $i$ of the resonant structure, $pe_i$ characterizes the electric filling factor in the region $i$, which is determined using a rigorous electromagnetic simulation based on the method of lines and verified with a finite element analysis technique[46,47] and defined in Eq.7. A Bragg reflector in this article is defined as a combination of dielectric and air in both radial and axial directions. The parameters $r_j$ and $h_j$ represent the dimensions of the structure in a particular region from 1 to n-reflector + 1 along the radial and axial direction respectively. The parameter $j + 2$ represents the metallic enclosure of the resonator. The total number of regions defines the number of reflectors (air+dielectric) plus the inner free-space region, for example, three for a single-Bragg-resonator and five for a double-Bragg-resonator.

$$pe = \frac{We_{material}}{We_{Total}} = \frac{\iiint_{V_{material}} \epsilon_{material} E.E^* dV}{\iiint_{V_{Total}} \epsilon(v) E.E^* dV} \quad (7)$$

In the case where the dielectric material permittivity is high (> 50), its thermal expansion would be the dominating factor in calculating the temperature coefficient of frequency and would result in the limiting factor to achieving small $\tau_f$. Alternatively, both the temperature coefficient of permittivity and the thermal expansion of the metal have the same significance if the dielectric permittivity is relatively low (< 10).

TABLE I: Coefficient of thermal expansion - CTE ($\alpha(T)$) and temperature coefficient of permittivity - TCP ($\tau\epsilon(T)$) fitting parameters from 2 to 350K and from 6 to 15GHz[14,49-51,53,57], as follow,

$$\alpha(T) \text{ or } \tau\epsilon(T) = a_0 T^0 + a_1 T^1 + a_2 T^2 + a_3 T^3$$

CTE - 2 to 30K

| $\alpha(T)$ (ppm/K) | $a_0$ | $a_1$ | $a_2$ | $a_3$ |
|---|---|---|---|---|
| $Al_2O_3 \perp$ | $6 \times 10^{-17}$ | $-2.7 \times 10^{-17}$ | $2.9 \times 10^{-18}$ | $8.9 \times 10^{-7}$ |
| $Al_2O_3$ " | $7.5 \times 10^{-11}$ | $-4 \times 10^{-11}$ | $5 \times 10^{-12}$ | $7.7 \times 10^{-7}$ |

CTE - 20 to 350K

| $\alpha(T)$ (ppm/K) | $a_0$ | $a_1$ | $a_2$ | $a_3$ |
|---|---|---|---|---|
| $Al_2O_3 \perp$ | 0.182 | -0.008 | $1.8 \times 10^{-4}$ | $-3.1 \times 10^{-7}$ |
| $Al_2O_3$ " | 0.080 | -0.009 | $1.7 \times 10^{-4}$ | $-2.9 \times 10^{-7}$ |

TCP

| $\tau\epsilon(T)$ | $a_0$ | $a_1$ | $a_2$ | $a_3$ |
|---|---|---|---|---|
| $Al_2O_3 \perp$ | -8.839 | 0.176 | $2.3 \times 10^{-3}$ | $-5.8 \times 10^{-6}$ |
| $Al_2O_3$ " | -25.314 | 0.582 | $1.8 \times 10^{-3}$ | $-6.7 \times 10^{-6}$ |

To make predictions, we relied on the data of dielectric materials properties from previously published data for sapphire and alumina[14,49-53] and of copper, brass and aluminum for the resonator enclosures[54,?-56]. From these data, we interpolated from polynomial fits to cover temperatures ranging from 4 to 350K and frequencies from 6 to 15GHz. The fit parameters are given in Table I and Table II, for the dielectrics and the metals respectively. For alumina, an isotropic material, its properties have been Frequency-Temperature sensitivity reduction with optimized microwave Bragg resonators estimated from sapphire perpendicular to crystal axis.

## 1. Single Bragg Reflector (SBR) Resonators

The single Bragg resonator (SBR), is composed of two discs of a high-quality monocrystalline sapphire, assembled together on top and bottom of a hollow cylindrical

TABLE II: Coefficient of Thermal Expansion (CTE) fitting parameters from 2 to 350K[54–56] as follow,

$$\alpha(T) = a_0 T^0 + a_1 T^1 + a_2 T^2 + a_3 T^3 + a_4 T^4 + a_5 T^5$$

| α(T) | $a_0$ | $a_1$ | $a_2$ | $a_3$ | $a_4$ | $a_5$ |
|---|---|---|---|---|---|---|
| Ag | -6.376 | 0.417 | $-3.0 \times 10^{-3}$ | $1.1 \times 10^{-5}$ | $2.1 \times 10^{-8}$ | $1.5 \times 10^{-11}$ |
| Al | -0.021 | 0.025 | $-3.7 \times 10^{-3}$ | $2.1 \times 10^{-4}$ | $3.6 \times 10^{-6}$ | $3.1 \times 10^{-8}$ |
| Au | -4.322 | 0.375 | $-3.3 \times 10^{-3}$ | $1.5 \times 10^{-5}$ | $3.2 \times 10^{-8}$ | $2.8 \times 10^{-11}$ |
| Cu | -6.250 | 0.410 | $-2.9 \times 10^{-3}$ | $1.1 \times 10^{-5}$ | $-1.9 \times 10^{-8}$ | $1.3 \times 10^{-11}$ |
| CuZn | -4.159 | 0.315 | $-1.7 \times 10^{-3}$ | $4.4 \times 10^{-6}$ | $-5.2 \times 10^{-9}$ | $2.3 \times 10^{-12}$ |
| Ti | -0.839 | 0.058 | $-9.5 \times 10^{-5}$ | $3.5 \times 10^{-8}$ | - | - |

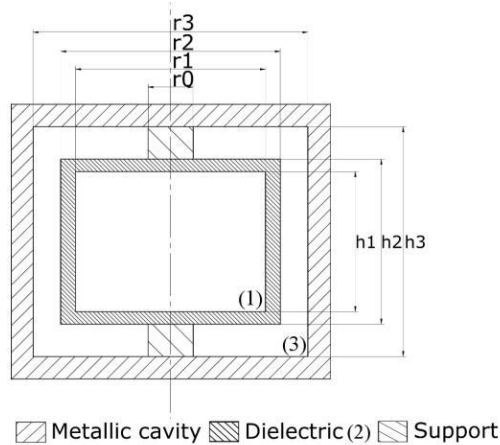

FIG. 2: Schematic of a single Bragg resonator design where $r_i$ correspond to the radii of the support, inner and outer dielectric layer, and the cavity respectively. $h_i$ represent the height of the inner and outer dielectric layer and the cavity, respectively. The white regions 1 and 3 are vacuum and finally the shaded region 2 is the dielectric.

Frequency-Temperature sensitivity reduction with optimized microwave Bragg resonators sapphire piece. It is then inserted into a metallic enclosure. We also define as per Eq.6, the different dielectric regions by their permittivity $\varepsilon_i$ with increasing numbers from inner to outer layers, denoted as $\varepsilon_1$, $\varepsilon_2$, and $\varepsilon_3$, respectively. Fig.2 introduces the parameters used to define a single Bragg resonator.

The operating mode of a Bragg resonator is the fundamental mode ($TE_{01\delta}$), which only has three components ($E_{phi}$, $H_r$, $H_z$), that means the radial and axial electric field does not exist, hence the radial and axial electric filling factors equal zero. The method of lines and FEM[46,47] software can compute the partial derivatives of frequency related to dimensions, by varying one dimension at a time and keeping all others to their fixed initial values.

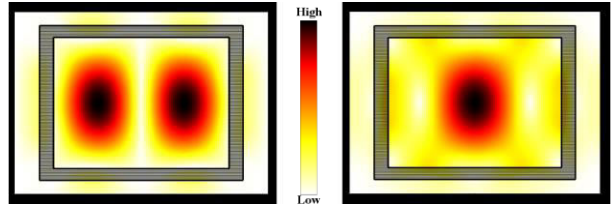

FIG. 3: Density plot of the single Bragg resonator, (left) represents the electric field and (right) the magnetic field. The shaded region illustrates the sapphire reflector. The color coding of the intensity field is darker while the field is maximum, whereas the color is very light when the intensity is small.

### 1.1. Sapphire reflector

The SBR dimensions are $r_1$ = 21.05mm, $r_2$ = 24.35mm, $h_1$ = 31mm, and $h_2$ = 36.6mm. The metallic enclosure is made of brass with $r_3$ = 30.45mm, and $h_3$ = 50.90mm. The support is made of copper to ensure a good thermal contact and keep the temperature of the sapphire stabilized, during measurements. Nitric acid is used to clean the sapphire reflector; the best Q-factor before the cleaning process was 170,000 and then after cleaning Q-factors as high as 241,000 were measured at 9.77GHz at room temperature. The fundamental Bragg mode ($TE_{01\delta}$) electric and magnetic field energy density plot distributions are shown in Fig.3. The resonance frequency of the mode depends on both the dimensions of the cavity and sapphire as well as its permittivity. Sapphire is an anisotropic material. It is then, required to write the $\kappa_{f\varepsilon_2}$ (Eq.6a)

Frequency-Temperature sensitivity reduction with optimized microwave Bragg resonators differently to take into account the thermal expansion coefficients of the dielectric and TCPs for perpendicular and parallel directions to the crystal axis (c-axis).

$$\kappa_{f\epsilon_2}\tau_{\epsilon_2} = \left(\frac{\partial f}{\partial \epsilon_{2\perp}}\frac{\epsilon_{2\perp}}{f}\right)_{\perp}\tau_{\epsilon_{2\perp}} + \left(\frac{\partial f}{\partial \epsilon_{2\|}}\frac{\epsilon_{2\|}}{f}\right)_{\|}\tau_{\epsilon_{2\|}} \qquad (8)$$

### 1.2. Alumina reflector

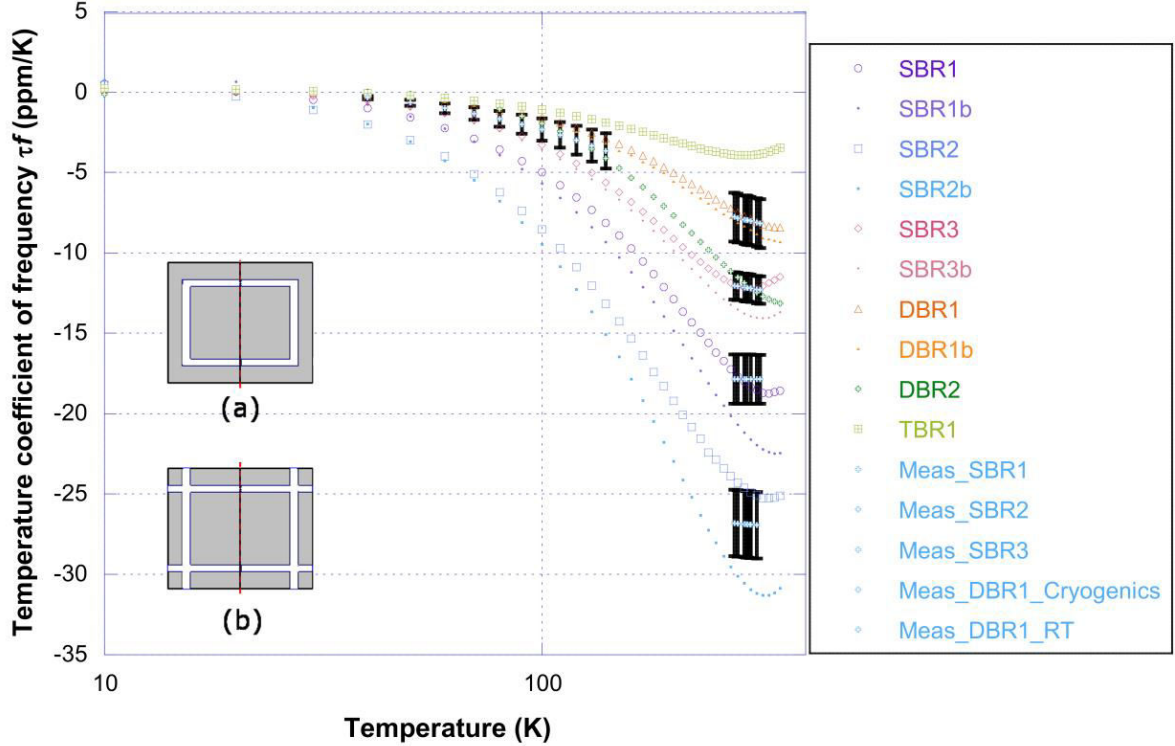

FIG. 4: Results of the predictions and measurements of single Bragg, and double Bragg resonators decomposed as follows: (SBR1) is a single sapphire Bragg resonator, (SBR2) is a single alumina Bragg resonator , (SBR3) is a single hybrid-Bragg-mode resonator, (DBR1) is a aapphire-alumina double Bragg resonator, (DBR2) is the prediction of the double Bragg resonator from Flory and Ko[32] and (TBR1) is the prediction of the triple Bragg resonator from Flory and Taber[31]. Inset (a) optimized design of Bragg resonator, inset (b) illustrates the "cross" in each corner of the resonator. This addition of dielectric material increases its contribution in the temperature coefficient of frequency.

In Fig.4, we report the $\tau_f$ prediction for a single alumina Bragg reflector which later served as the second reflector of a sapphire-alumina double Bragg resonator. This alumina reflector is made of a very high purity material which exhibits a loss tangent of about $2.4 \times 10^{-6} f[GHz]$[43], resonating at 7.20GHz with an unloaded Q-factor of 97,300 at Frequency-Temperature sensitivity reduction with optimized microwave Bragg resonators room temperature. Sapphire and alumina are based on the same crystalline structure $Al_2O_3$. Alumina is an isotropic ceramic material, therefore we used the data from the sapphire perpendicular permittivity to scale the TCP dependence to room temperature

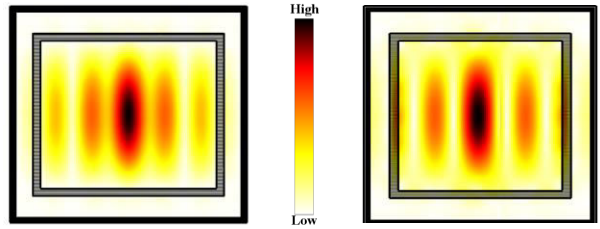

FIG. 5: Energy density plots for the hybrid-Bragg-mode resonator, (left) represents the electric field and (right) the magnetic field. The shaded region illustrates the sapphire reflector. The color coding of the intensity field is darker where the field is maximum, whereas the color is very light where the intensity is small.

resulting from the alumina permittivity (9.73)[22,43].

**1.3. Hybrid-Bragg-mode resonator**

The hybrid-Bragg-mode combines Bragg and whispering gallery-like modes (see Fig.5). It has six field components of which three are dominant. Its polarization is similar to a transverse electric mode, excited with magnetic loop probes along the z-axis. The perpendicular component of this mode meets the boundary conditions for allowing Bragg reflections. More particularly, the hybrid-Bragg-mode resonator is still considered as a single-Bragg-resonator operating on a hybrid electromagnetic mode (WGE) but exhibits the same properties and boundary conditions along the azimuthal direction as the $TE_{0,1,\delta}$ Bragg resonators. The dimensions of the alumina are similar as the single Alumina Bragg resonator. The resonance frequency of a quasi-TE mode with an azimuthal variation of one is about 13.41GHz with an unloaded Q-factor of about 191,000 at room temperature[43]. This results corresponds to a factor 6 to 10 times above the loss tangent limit of the alumina. With this Q-factor enhancement mechanism, Bragg resonators may be made of materials other than ultra-low-loss crystals[42]. This has been demonstrated with this alumina resonator achieving a result as good as a sapphire whispering gallery mode resonator. The results are reported in Fig4.

*2. Double Bragg resonator (DBR*

Frequency-Temperature sensitivity reduction with optimized microwave Bragg resonators

The double Bragg resonator is composed of an inner and an outer reflector made of sapphire and alumina materials. The DBR here has been scaled from our previous modeling[41,47] for which the schematic and dimensions are given in Fig.6. Using Eq.6, with *i* and *j* equal to 5 and 2 respectively, we can determine the $\tau_f$ of a double Bragg resonator (DBR). The electric energy density plot of this double-Bragg resonator is given in Fig.7. It shows how the field is very well confined in the inner free-space region. With such 359 a structure, it is expected to achieve a Q-factor of about 500,000 at 9.79GHz at room temperature. However, we have been unable to couple well enough to reach such a high-Q factor. Thus, we drilled two small probe holes in the second reflector made of alu- mina to couple more strongly to the axial magnetic field components ($H_z$). This maneuver was unsuccessful. We then used a cryo-refrigerator with a LakeShore 332 temperature controller to monitor the frequency and Q-factor at given temperatures. Decreasing in temperature helped move away spurious dielectric modes, which damp the Q-factor. They are mostly located in the horizontal reflectors. We employed this technique to predict the $\tau_f$ value in a similar way to the SBR case but at cryogenic temperatures (see Fig.4). The extra reflector is made with one hollow cylinder and two plates of alumina ceramic.

## B. Comparison with other resonators

For oscillator applications, Bragg resonators are designed and expected to perform much better than whispering gallery mode resonators as their expected Q-factor may reach one million[31,38]. If for a Bragg reflector the central reflector is much more important in terms of Q-factor, it is the same in terms of temperature coefficient. That means for a Bragg resonator the second reflector makes less contribution than the first to the frequency-temperature dependence. However, room temperature oscillators don't have an inversion frequency-temperature point to operate at, therefore the coefficient of frequency to temperature ($\tau_f$) dependence is of extreme importance to either find a temperature which has a $\tau_f$ equals to zero or use a temperature-compensated design. It might be worth considering a material for the second layer with an opposite sign temperature coefficient with a value is much larger than the material of the first reflector. This should enable compensation of the frequency-temperature coefficient of the whole resonator as

Frequency-Temperature sensitivity reduction with optimized microwave Bragg resonators shown in[35,58] or by using a gas buffer[59]. In this work, we only want to characterize how $\tau_f$ varies depending on the material and design we choose, in particular the hybrid Bragg mode. Whispering gallery mode resonators are oftenly used even at room temperature[3,4], but their $\tau_f$ varies from -50 to -70ppm/K depending on the chosen polarization (WGE or WGH) respectively[22,60]. Previously designed Bragg resonators[31,32] (see Fig.8) have never been characterized in terms of $\tau_f$. We made a prediction following the same approach as described in Section III.A. We have been able to compare their sensitivity with our new modeled Bragg resonators as well as the hybrid Bragg mode resonators, for which results have been reported in Fig.4. From these results, it is clear that any Bragg resonator has a better $\tau_f$ than a WGM resonator with similar Q-factor and frequency band. We also investigated the change of the metallic enclosure to measure its effect on the $\tau_f$ coefficient. It has shown the influence of the metallic enclosure due to the choice of the metals given in TableII is at its highest difference at room temperature within 5% of the reported values in Fig.4 in the case of a single Bragg resonator. This falls within the uncertainty evaluation of $\tau_f$.

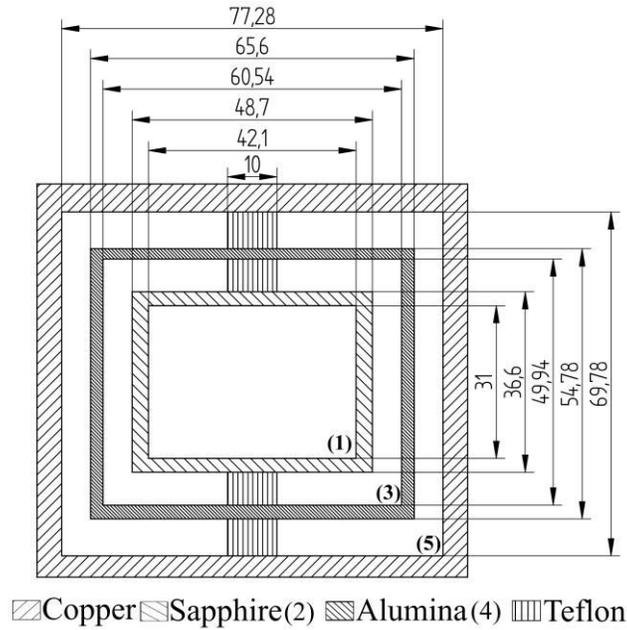

FIG. 6: Schematic and dimensions of the double Bragg Resonator (DBR) made of sapphire and alumina reflectors enclosed and centered in a copper cavity. The different regions are numbered from 1 to 5, where 1, 3, and 5 are free-space, 2 is sapphire and 4 is alumina.

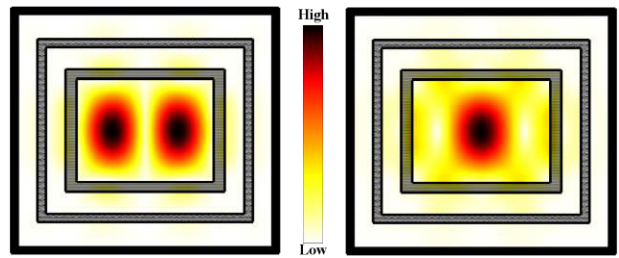

FIG. 7: Energy density plots for the Double Bragg resonator, (left) represents the electric field and (right) the magnetic field. The dark inner shaded region illustrates the sapphire reflector and the light outer shaded one the alumina reflector. The color coding of the intensity field is darker where the field is maximum, whereas the color is very light where the intensity is small.

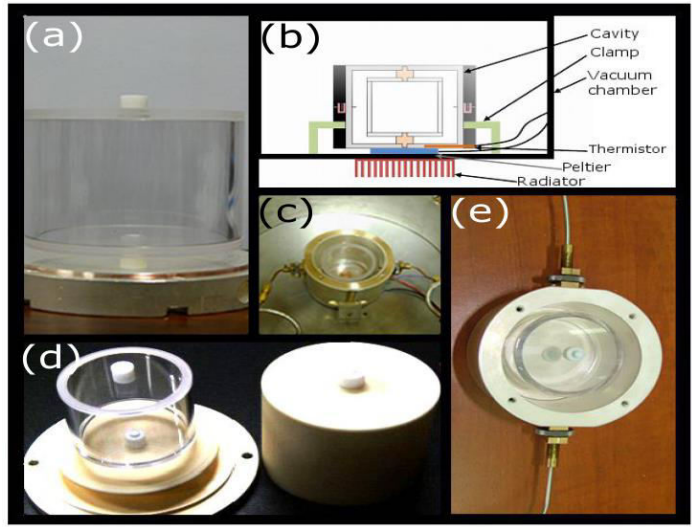

FIG. 8: Illustrations of our Single Bragg, Double Bragg resonators including the experimental and schematic of the setup, as follows: (a) a Single Sapphire Bragg resonator, (b) the schematic of the setup, (c) the photograph of the mounted cavity into the vacuum chamber, (d) the Sapphire-Alumina Double Bragg resonator, and (e) is the top view of the Single Bragg resonator inside the copper cavity.

## IV. DISCUSSION

We presented and compared different topologies of Bragg resonators. The $\tau_f$ of single

Frequency-Temperature sensitivity reduction with optimized microwave Bragg resonators sapphire Bragg reflector resonator is already about two to three times lower than any WGM resonators with similar Q-factors (∼ 200, 000). Optimized distributed Bragg resonators from our new modeling Fig.4(a) can already offer a 10% increase in Q-factor, a 30% more compact structure and reduce the sensitivity to temperature. Also, we demonstrated that the use of a cheap material (lossy at microwave and with large TCP value) could be greatly enhanced by using hybrid-Bragg-mode achieving a Q-factor as high as WGM resonators. Moreover, by implementing the hybrid-Bragg-mode a Q-factor enhancement that could reach a value ten times higher than the loss tangent limit, while downsizing the volume by about 30% and reducing $\tau_f$ by a factor of 6. The increasing confinement from 80% to 90% in the free-space region of the resonator is the result of the whispering gallery-like mode propagating along $r$ and $\theta$-directions due to the Bragg boundary conditions. This way less field is concentrated in the dielectric of the reflectors, which benefit the Q-factor and $\tau_f$. It shows that the hybrid-Bragg-mode is less sensitive to temperature than any other types of Bragg resonator made of low-loss material like sapphire, such as single and double sapphire Bragg resonators. Simulations and experiments showed that the Q-factor gain in multi-layered Bragg resonators reach a limit with three reflectors in both radial and axial variations. After that the increase is not so relevant compared to the resonator size. In terms of the electric energy confined in the reflector, a single Bragg resonator has 23% in the dielectric, a double-layer has 1.8% in the outer dielectric layer and a three-layer has 0.05% energy confined in the last dielectric layer of its reflector. For the same reason as the hybrid Bragg mode, the more we confine energy in the free space the less field is in the reflectors, thus the less there is in the dielectric which makes the outer reflectors even less sensitive to temperature variations and at the same time serves as a shield for the resonator. Depending on the applications, size and weight may be of a serious issue, therefore alternatives may be sought. Temperature compensation techniques with dual opposite sign CTE materials or other types of modes could be used. All factors of improvement shown in this paper, make Bragg resonators operating on a hybrid Bragg mode very promising in the design of new types of filters, sensors and oscillators at room temperature.

**ACKNOWLEDGMENTS**

This research is jointly supported under the Australian Research Council funding scheme: Discovery Project (DP160100253). The authors would also like to thank the

Frequency-Temperature sensitivity reduction with optimized microwave Bragg resonators

French Research Agency (CNRS), Labex Sigma-Lim (No. ANR-10-LABX-0074-01) and the French-Australian Science and Technology Program: International Science Linkages ISL-DSIIR, Project number FR100027. The authors also thank le Conseil Régional du Limousin, cluster de calcul CALI (Calcul en Limousin). Finally, we thank The University of Western Australia - Advanced Science Program.